# Provocative radio transients and base rate bias: a Bayesian argument for conservatism

________________________________________________________________

Thomas W. Hair, Ph.D.

*College of Arts and Sciences, Florida Gulf Coast University, 10501 FGCU Blvd, South, Fort Myers, FL 33965, USA*
*Email:twhair@fgcu.edu*

**Abstract:** Most searches for alien radio transmission have focused on finding omni-directional or purposefully earth-directed beams of enduring duration. However, most of the interesting signals so far detected have been transient and non-repeatable in nature. These signals could very well be the first data points in an ever-growing data base of such signals used to construct a probabilistic argument for the existence of extraterrestrial intelligence. This paper looks at the effect base rate bias could have on deciding which signals to include in such an archive based upon the likely assumption that our ability to discern natural from artificial signals will be less than perfect.

**Motivation**

Imagine one thousand years from now a vibrant human presence throughout the solar system…from the balmy shores of a warm, wet Earth to the frigid, arid plains of Mars to even more remote outposts like Ceres, Ganymede, and Titan…and of a hundred interplanetary ships that link them for trade and travel. Take this vision further, say another ten thousand years more, and imagine the solar system, and quite possibly some nearby solar systems as well, completely subsumed in every meaningful way by a human presence [7]. If humanity is to continue to persist for any significant time into the future, then exploitation of the immense natural resources of our home solar system would seem to be a foregone necessity.

Nothing more than the aforementioned conservative scenario is required to detect an extra-terrestrial intelligence within several thousand light-years from Earth…no vast and supremely intelligent, galaxy spanning beings need apply. A humble group of dedicated extra-terrestrial survivalists is all that is required. Naysayers may point out that as we, as a civilization, go digital and line of sight in our modes of communication that the window for detection of any nascent alien civilization will be very short and thus the probability of detecting one of them will be very low given the immense temporal spread that the age of the universe implies between us and any number of them [5].

We may very well be going radio *communications* quiet here on the surface of the Earth [3], but this improbable thesis of a radio quiet Earth fails the further we get from our primordial gravity well. At present, the powerful parabolic radar dish at Arecibo, Puerto Rico is used to track asteroids like Vesta tens of millions of miles from home and the phased array Cobra Dane radar on Shemya Island, Alaska and at least a half dozen similar devices hammer megawatts of radio energy into deep space on a continual basis to track everything from potentially lethal near Earth objects and incoming enemy intercontinental ballistic missiles to wayward wrenches dropped by space shuttle astronauts in the 1980s.

**Searching for Extraterrestrial Radio Transients**

If humanity is to create a viable and complex solar society and a necessarily equally viable and complex inter-nodal, intra-stellar communications and control network in the coming millennia, then how do we exploit this idea of a randomly noisy human generated solar civilization to find another intelligence in the greater void? As pointed out by the author and others [1], *should we observe such activities by ETI, the signals would appear to us as transient events. But such quick, powerful bursts would only be verifiable by a 'staring' strategy, with smaller dishes looking continuously at the skies, most profitably at the galactic plane. Once such a burst appears, watching time can be focused on such possible sites, perhaps with some dishes linked so their effect could be coherent, raising the detection capacity of the network.* This statement suggests that a new strategy of scanning the sky for powerful directed signals is advantageous, and that once a coherent signal is detected that a strategy of long-term monitoring of that region of the sky be undertaken to detect more transient events at the same and other frequencies to build a probabilistic picture.

Throughout SETI program history several provocative transients have been detected, almost all of which never repeat. The WOW! signal being the most famous of these. Other less famous examples include Sullivan, et al. [10], which "recorded intriguing, non-repeatable, narrowband signals, apparently not of manmade origin and with some degree of concentration toward the galactic plane…" and similar searches which detected one-time signals that were not repeated [8,4,11]. Very few of these searches lasted more than an hour. But what if, as mentioned previously, a strategy of long duration monitoring of a fixed portion of the sky were initiated, how would we separate the enormous amount of natural transients from their artificial counterparts using an imperfect detection algorithm?

A somewhat recent example of this conundrum is the radio transient GCRT J1745-3009. We simply don't know what it is. Many reasonable natural explanations have been proposed and are currently undergoing peer review [12,9], but the question remains as to how does one definitively say that a seemingly coherent, narrow band radio transient is natural or not. Given the possible kilo-parsec distances from which this signal may emanate and the natural forces that distort it along a path to us, the answer is not certain and demands that we apply a genuine conservatism when building a probabilistic case for extraterrestrial intelligence based upon the collection of large numbers of provocative radio transients.

**Base Rate Bias Considerations**

Base rate bias, also called base rate neglect or the base rate fallacy, is an error that occurs when the conditional probability of some hypothesis H given some evidence E is assessed without taking into account the prior probability or *base rate* of H and the total probability of evidence E [2]. It happens when the values of sensitivity and specificity (which only depend on the test itself) are used in place of positive predictive value and negative predictive value (which depend on both the test and the baseline prevalence of the event).

Base rate bias is also one of the cornerstones Bayesian statistics, as it stems directly from Bayes' famous theorem

$$P(A|B) = \frac{P(A) \cdot P(B|A)}{P(B)}$$

Expanding the probability $P(B)$ for the set of all $n$ possible, mutually exclusive outcomes $A$ we arrive at

$$P(B) = \sum_{i=1}^{n} P(A_i) \cdot P(B|A_i)$$

Combining the two equations yields a generally more useful statement of Bayes' Theorem

$$P(A|B) = \frac{P(A) \cdot P(B|A)}{\sum_{i=1}^{n} P(A_i) \cdot P(B|A_i)}$$

with the denominator $\sum_{i=1}^{n} P(A_i) \cdot P(B|A_i)$ being defined as the Law of Total Probability or Total Alternatives, where the summation can be interpreted as a *weighted* average and the marginal probability $P(B)$ interpreted as the *average* probability [13].

As a thought experiment let us suppose, for example, in a dedicated, wide-spectrum search of the galactic center where 90% of the galaxy's stars lie within just 9% of the sky from our perspective, a one year *long stare* search endeavor yields 1 million radio transient signals of interest. If, unknown to us, there are 10 actual artificial (i.e., extraterrestrial) signals and 999,990 natural signals, then the base rate probability of one random signal from the search being artificial is thus 0.00001 and the base rate probability of a random signal being a natural source is

0.99999. In an attempt to find these artificial signals within the much greater natural ones a complex algorithm must *recognize* their artificialness. In this example, we can imagine an algorithm that has two failure rates of 0.001:

1. If the algorithm recognizes an artificial signal, it will correctly identify it with a probability of 0.999, and mistakenly fail to recognize it with probability 0.001 (in other words, the false-negative rate is 0.1%).
2. If the algorithm recognizes a natural signal, it will correctly identify it with a probability of 0.999, but it will mistakenly misidentify the signal as artificial with a probability of 0.001 (the false-positive rate is 0.1%).

So, the failure rate of the algorithmic based transient detection system is always 0.1% in this example.

Now suppose a radio transient is received and recognized by the algorithm as being of intelligent extraterrestrial origin. What is the chance it actually is artificial? Someone exhibiting base rate bias would incorrectly claim that there is a 99.9% chance that this signal is extra-terrestrial and pop the champagne, because the failure rate of the algorithm is always a measly 0.1%. Although this seems to make sense, it is actually faulty reasoning. The application of Bayes' Theorem below shows that the chance the signal is extraterrestrial is actually near 1%, not near 99.9%.

Let *P(ET|ID)* be the probability of correctly identifying a transient signal as being artificial, that is, the signal is actually artificial given the algorithm identifies it as such and where *P(ET)* is the base rate probability that any given transient received is artificial, *P(ID)* is the probability a transient is determined by the algorithm to be artificial, and *P(ID|ET)* is the probability the algorithm identifies a transient as artificial given that it is indeed artificial. Therefore, using Bayes' Theorem

$$P(ET|ID) = \frac{P(ID|ET) \times P(ET)}{P(ID)}$$

can be re-written as

$$= \frac{P(ID|ET) \times P(ET)}{P(ID|ET) \times P(ET) + P(ID|\sim ET) \times P(\sim ET)}$$

with $P(\sim ET)$ and $P(ID|\sim ET)$ being complements of their aforementioned probabilities and the denominator $P(ID|ET) \times P(ET) + P(ID|\sim ET) \times P(\sim ET)$ expressing the Law of Total Probability's two alternatives of correctly or incorrectly identifying a transient signal as artificial. Finally, inserting the numbers used in the previous discussion yields

$$= \frac{0.999 \times 0.00001}{(0.999 \times 0.00001) + (0.001 \times 0.99999)} = 0.0099 \approx 1\%$$

So, the actual probability that a signal recognized by the detection algorithm as artificial is, in fact artificial, computes to a very low one percent. The fallacy arises from confusing two distinctly different failure rates. The first being the number of natural signals per algorithm recognition and the second being the number of non-recognitions per artificial signal received. They are wholly unrelated quantities, and there is no reason one has to equal the other. They don't even have to be roughly equal.

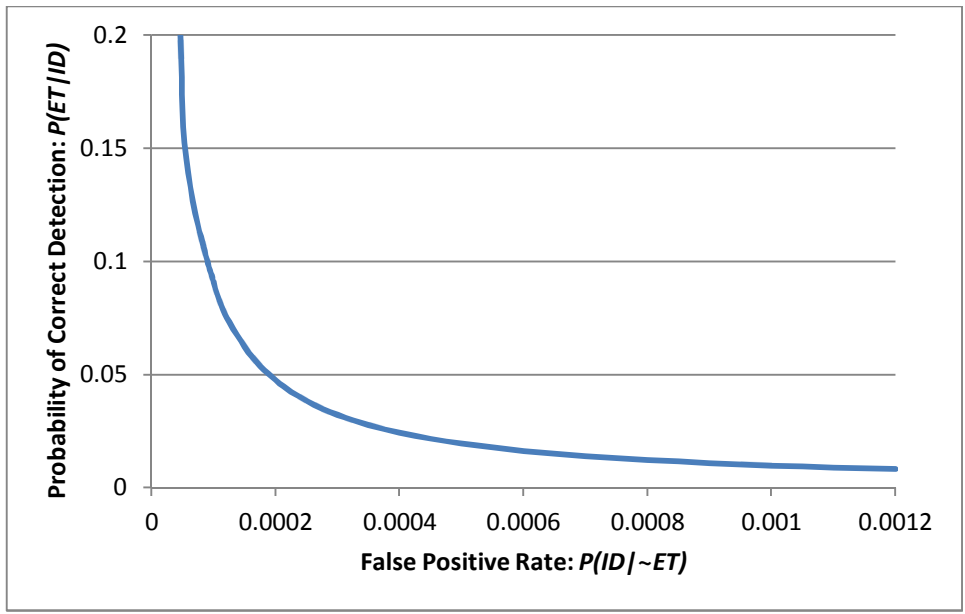

**Figure 1:** Graphical comparison showing the correct detection of an artificial extraterrestrial signal $P(ET|ID)$ as a function of the false positive rate $P(ID|\sim ET)$ highlighting the need for extreme algorithmic accuracy.

Going back to the original thought experiment, one can compute the rate at which natural signals are being algorithmically misidentified as artificial. Imagine that the search's entire population of one million radio transients pass through the algorithm. All ten of the artificial signals will be recognized correctly by the algorithm (actually 9.99 of them!), but so will about 1000 of the 999,990 natural signals (again, actually 999.99). Therefore, about 1,010 transient signals will be recognized as artificial by the algorithm, among which only about 10 will actually be artificial (see Table 1). The base rate bias in this example is primarily fallacious because there are many more natural signals than artificial ones. If the algorithm were checking approximately as many artificial signals as natural ones, and the false-positive rate and the false-negative rate were nearly equal, then the probability of mis-identification would be about the same as the false-positive rate of the algorithm.

| Radio Transients Received | |
|---|---|
| 10 artificial transients | 999,990 natural transients |
| Algorithm Detections (false positive rate = 0.01) | |
| 10 recognized as artificial | 1000 recognized as artificial |

**Table 1:** A tabular representation of the base rate bias problem. The algorithm will detect all the artificial transients, but because it also has a false positive rate of 0.001 it will also incorrectly identify 1000 natural signals as artificial. Thus, the actual probability a transient identified as artificial, is indeed artificial, is about 0.01.

Figure 1 shows that the false positive rate $P(ID|\sim ET)$ would have to be significantly lowered for the probability of a correct detection $P(ET|ID)$ to rise above even 10%. To reach a threshold of 90% the algorithm would need a fidelity three orders of magnitude better (i.e., 1000 times more sensitivity) than the example provided; something that cannot be effectively shown in the figure. However, using a false positive rate of $1 \times 10^6$ and applying it to our formula from the first example yields a correct detection rate of

$$= \frac{0.999999 \times 0.00001}{(0.999999 \times 0.00001) + (0.000001 \times 0.99999)} = 0.917 \approx 92\%$$

**A Poisson Approximation to a Binomial Scenario**

The large scale sifting of these provocative radio transients in the search for ones emanating from extraterrestrial civilizations can be thought of as a series of binomial trials where the probability of $x$ successfully recognized artificial signals collected within $n$ trials with probability of success $p$ on each trial is

$$binom(n, p, x) = nCx \cdot p^x \cdot q^{n-x}$$

with $q = 1 - p$ and $nCx$ representing the binomial coefficient

$$nCx = \frac{n!}{(n-x)!\, x!}$$

where $n$ represents the total number of signals collected [13]. If no valid artificial signal is correctly detected in $n$ independent trials, then $x$ is zero. The probability of at least one valid signal being detected then becomes $1 - binom(n, p, 0)$ for all values of both $n$ and $p$. However, when $n$ is large and $p$ is small, as is the case in the radio transient search presented, binomial probabilities are often approximated by means of the Poisson distribution with

$$f(x;\lambda) = \frac{\lambda^x e^{-\lambda}}{x!} \quad for\ x = 0,1,2,\ldots \quad \lambda > 0$$

and $\lambda$ equal to the product *np*. The measurement of these radio transients can thus be thought of as a random physical process that is in part controlled by some sort of chance mechanism (i.e., the detection algorithm). What characterizes such a Poisson process is its time dependence, namely, the fact that certain events do or do not take place (depending on chance) at regular intervals of time [6].

Therefore, in order to find the probability of *x* successful detections during a time interval of length *T*, we divide the interval into *n* equal parts of length $\Delta t$, so that $T = n \cdot \Delta t$, and make the following assumptions:

1. The probability of a success during a very small time interval $\Delta t$ is given by $\alpha \cdot \Delta t$.
2. The probability of more than one success during such a small time interval $\Delta t$ is negligible.
3. The probability of a success during such a time interval does not depend on what happened prior to that time.

This means that the assumptions underlying the binomial distribution are satisfied, and the probability of x successes in the time interval T is given by the binomial probability

$$binom(n,p,x) \quad with \quad n = \frac{T}{\Delta t} \quad and \quad p = \alpha \cdot \Delta t$$

Lastly, we find that when $n \to \infty$ the probability of x successes during the time interval T is given by the corresponding Poisson probability with the parameter

$$\lambda = n \cdot p = \frac{T}{\Delta t} \cdot (\alpha \cdot \Delta t) = \alpha \cdot T$$

Since $\lambda$ is the mean of this Poisson distribution it should be noted that $\alpha$ is the average number of successes per time unit [6].

Returning to the original thought experiment, if sometime in the not too distant future a large radio telescope like the upcoming Square Kilometer Array (SKA) is set to the task of a *long stare* as previously described it would take, for example, $T = 2.74$ years to collect 1,000,000 provocative radio transients at 1,000 per $\Delta t = 1$ day. If it is assumed there are 10 artificial signals within the total radio transients (something that would not be known), then the probability any given radio transient is artificial is $p = 0.0001$. In fact, the mean number of radio transients that must be detected and processed by the algorithm until the first artificial one is received follows a geometric distribution given by

$$g(x;p) = p(1-p)^{x-1} \quad for\ x = 1,2,3\ldots$$

where the mean is

$$\mu = 1/p = 100{,}000 \text{ radio transients.}$$

In this scenario then it would not be until the 100$^{th}$ day of observation, on average, into a 2.74 year survey that the first artificial radio transient is received.

**Conclusions**

As SETI techniques begin to turn away from the search for purposefully Earth-directed signals of enduring duration and more toward finding the ephemeral leakage associated with the complex business of maintaining an interstellar civilization, large or small, then we must come to terms with the fact that most if not all of our detections will be one of a kind…never to repeat with exactly the same parametric characteristics or in the same position in the sky. With this in mind it becomes particularly important to collect a large data set of provocative (i.e., possibly indicative of extraterrestrial intelligence) radio transients through a *long stare* strategy in order to build up a compelling case that they cannot all be explained by natural phenomena in much the same way the dedicated scientists at CERN used their diligence and perseverance to recently claim the Higgs Boson exists to an extremely high degree of confidence. In consideration of this it is important to realize that the finest mathematical sieve will always gather up far more straw from the haystack than it does those very special needles.

**References**

[1]    Benford, J. & Hair, T. (2010) *SETI and Detectability*, www.centauri-dreams.org/?p=13340.

[2]    Bar-Hillel, M. (1980). *Acta Psychologica* **44**, 211.

[3]    Forgan, D. & Nichol, R. (2010) *International Journal of Astrobiology* **10** (2), 77.

[4]    Gray, R. & Marvel, K. (2001). *The Astrophysical Journal* **546** (2): 1171.

[5]    Hair, T. (2011). *International Journal of Astrobiology* **10**, 131.

[6]    Johnson, R. (2005). *Miller & Freund's Probability and Statistics for Engineers, 7$^{th}$ Edition*, Prentice Hall.

[7]    Kardashev, N.S. (1964). *Soviet Astronomy* **8**, 217.

[8]    Shostak, S. & Tarter, J. (1985). *Acta Astronautica* **12**, 369.

[9]     Spreeuw, H.; Scheers, B.; Braun, R.; Wijers, R.;Miller-Jones, J.; Stappers, B.; Fender, R. (2009). *Astronomy and Astrophysics* **502** (2), 549.

[10]    Sullivan, W.T. III; Wellington K.; Shostak, S.;Backus, P.; Cordes, J. (1997), *International Astronomical Union Colloquium* **161**, 653.

[11]    Tarter, J. (2001). *Annual Review of Astronomy and Astrophysics* **39**, 511.

[12]    Zhu, W. & Xu, R. (2006). *Monthly Notices of the Royal Astronomical Society: Letters* **365**, 16.

[13]    Zwillinger, D. & Kokoska, S. (2000) *CRC Standard Probability and Statistics Tables and Formulae*, CRC Press.